\documentclass[aps,prd,preprint,preprintnumbers,superscriptaddress,showpacs]{revtex4-1}
\usepackage{graphicx}

\usepackage{ulem}
\usepackage{color}
\definecolor{My_red}        {cmyk}{0.00,1.00,1.00,0.20}

\newcommand{\bmat}{\left(\begin{array}}
\newcommand{\emat}{\end{array}\right)}
\newcommand{\beq}{\begin{equation}}
\newcommand{\eeq}{\end{equation}}

\def\bwt{\begin{widetext}}
\def\ewt{\end{widetext}}
\def\be{\begin{equation}}
\def\ee{\end{equation}}
\def\bea{\begin{eqnarray}}
\def\eea{\end{eqnarray}}
\def\bean{\begin{eqnarray*}}
\def\eean{\end{eqnarray*}}
\def\bary{\begin{array}}
\def\eary{\end{array}}
\def\bit{\begin{itemize}}
\def\eit{\end{itemize}}

\def\su5u1{SU(5) \times U(1)}
\def\fsu5u1{SU(5) \times U(1)'}
\def\so10{SO(10)}
\def\sq20{SO(10) \times SO(10)}

\def\bwt{\begin{widetext}}
\def\ewt{\end{widetext}}
\def\be{\begin{equation}}
\def\ee{\end{equation}}
\def\bea{\begin{eqnarray}}
\def\eea{\end{eqnarray}}
\def\bean{\begin{eqnarray*}}
\def\eean{\end{eqnarray*}}
\def\bary{\begin{array}}
\def\eary{\end{array}}
\def\bit{\begin{itemize}}
\def\eit{\end{itemize}}

\def\su5u1{SU(5) \times U(1)}
\def\fsu5u1{SU(5) \times U(1)'}
\def\so10{SO(10)}
\def\sq20{SO(10) \times SO(10)}

\usepackage[centertags]{amsmath}
\usepackage{amssymb}
\newcommand{\Z}{{\mathbb Z}}

\begin{document}

\title{The $750$~GeV Diphoton Excesses in a Realistic D-brane Model}

\author{Tianjun Li}

\affiliation{State Key Laboratory of Theoretical Physics and 
Kavli Institute for Theoretical Physics China (KITPC),
Institute of Theoretical Physics, Chinese Academy of Sciences, 
Beijing 100190, P. R. China}

\affiliation{
School of Physical Electronics, University of Electronic Science and Technology of China, 
Chengdu 610054, P. R. China 
}

\author{James A. Maxin}

\affiliation{Department of Physics and Engineering Physics,
  The University of Tulsa, Tulsa, OK 74104, USA}

\author{Van E. Mayes}

\affiliation{Department of Physics, University of Houston-Clear Lake,
  Houston, TX 77058, USA}

\author{Dimitri V. Nanopoulos}

\affiliation{George P. and Cynthia W. Mitchell Institute for Fundamental Physics
  and Astronomy, Texas A$\&$M University, College Station, TX 77843, USA}

\affiliation{Astroparticle Physics Group, Houston Advanced Research Center (HARC),
  Mitchell Campus, Woodlands, TX 77381, USA}

\affiliation{Academy of Athens, Division of Natural Sciences, 28 Panepistimiou Avenue,
  Athens 10679, Greece}

\date{\today}

\begin{abstract}

We study the diphoton excesses near $750$~GeV recently reported by the ATLAS and CMS
collaborations within the context of a phenomenologically interesting 
intersecting/magnetized D-brane model on a toroidal orientifold.  
It is shown that the model contains a SM singlet scalar as well as vector-like
quarks and leptons.  In addition, it is shown that the singlet scalar has Yukawa couplings
with vector-like quarks and leptons such that it may be produced in proton-proton
collisions via gluon fusion as well as decay to diphotons through loops
involving the vector-like quarks.  Moreover, the required vector-like quarks and leptons
may appear in complete $SU(5)$ multiplets so that gauge coupling unification may be 
maintained.  Finally, it is shown that the diphoton signal may be accommodated within 
the model.  

\end{abstract}


\preprint{ACT-03-16, MI-TH-1611}

\maketitle

\section{Introduction}

Recently, the ATLAS~\cite{bib:ATLAS_diphoton} and CMS~\cite{bib:CMS_diphoton} collaborations 
have both reported an excess in the diphoton channel near $750$~GeV.
With an integrated luminosity of 3.2 ${\rm fb}^{-1}$, the ATLAS collaboration has observed a local $3.6\sigma$
excess at a diphoton invariant mass of around 747~GeV, assuming
a narrow width resonance. For a wider width resonance, the signal significance
increases to $3.9\sigma$ with a preferred width of about 45~GeV.
With an integrated luminosity of 2.6 ${\rm fb}^{-1}$, the CMS collaboration has also observed a diphoton excess
with a local significance of $2.6\sigma$ at invariant mass of around 760 GeV. Assuming
a decay width of around 45~GeV, the significance reduces to $2\sigma$ in this case.
The corresponding excesses in the cross section can be roughly estimated as
$\sigma_{pp\to \gamma \gamma}^{13~ {\rm TeV}} \sim 3-13~{\rm fb}$~~\cite{bib:ATLAS_diphoton, bib:CMS_diphoton}.
  While this is well below the threshold to claim a discovery, this excess could be the first signal of physics
beyond the Standard Model.  As such, it is worthwhile to consider possible models of new physics which may explain 
the excess.  Indeed, many groups have proposed such possible 
explanations~\cite{Dutta:2015wqh,Falkowski,diphoton-first,diphoton-rest,Hall:2015xds,Patel:2015ulo,Ding:2015rxx,Allanach:2015ixl,Aydemir:2016qqj,Ding:2016udc,Han:2016bus,Dutta:2016jqn,Li:2016xcj,Hamada:2016vwk,Staub:2016dxq,Baek:2016uqf,Ko:2016sxg,Domingo:2016unq,Cvetic:2016omj,Ren:2016gyg,Lazarides:2016ofd}

Perhaps the simplest explanation for the excess is the addition of a SM singlet scalar with a mass
near $750$~GeV along with additional vector-like multiplets of colored particles.
With this set-up, the singlet may be produced via gluon fusion with the vector-like particles
appearing in loops.  Similarly, the singlet may decay to diphotons. 
However, in order to preserve gauge coupling unification as in supersymmetric versions of the SM, 
these vector-like states should come in complete multiplets of $SU(5)$.  Moreover, 
to preserve unification and avoid Landau poles, the types and numbers of $SU(5)$ 
multiplets is restricted~\cite{Dutta:2016jqn}.

Such light vector-like multiplets are often found in models constructed within the framework
of string theory\cite{Dienes:1996du}.  
Indeed, vector-like states are generically present in intersecting/magnetized
D-brane models on orientifold backgrounds~\cite{Berkooz:1996km,
Ibanez:2001nd, Blumenhagen:2001te, CSU, Cvetic:2002pj, Cvetic:2004ui, Cvetic:2004nk, 
Cvetic:2005bn, Chen:2005ab, Chen:2005mj, Blumenhagen:2005mu,Chen:2006gd,Chen:2006ip}. 
One such model satisfying all global consistency conditions has been constructed
from intersecting/magnetized D-branes within the context 
of Type II orientifold compactifications~\cite{Chen:2007zu,Chen:2007px} 
on a $T^6/(\Z_2 \times \Z_2)$ 
background.  
This model corresponds to the MSSM with  three generations 
of quarks and leptons as well as a single pair of Higgs 
fields.  
The model contains a minimal amount of exotic matter, which may be decoupled
from the low-energy sector.  In addition, the tree-level gauge couplings are automatically unified
at the string scale~\cite{Chen:2007zu,Chen:2007px}. Finally, the Yukawa couplings are allowed 
by global $U(1)$ symmetries, and it is possible to obtain correct masses and mixings 
for quarks and charged leptons.  Thus, this is a phenomenologically
interesting model worthy of detailed study.   

In the following, we briefly summarize the intersecting D-brane model
under study, which is a variation of the model discussed above.  
It is shown that vector-like quarks are present in the model, 
and that these states may appear in complete multiplets of $SU(5)$ so that 
gauge coupling unification may be maintained. Furthermore, it is shown
that there are SM singlets in the model which have Yukawa couplings to 
the vector-like quarks and leptons. It should be emphasized that this is a 
non-trivial result as these fields carry global $U(1)$ charges, under which
the Yukawa coupling must be neutral.  Finally, we show that the diphoton
excesses can be accomodated in the model.

\section{The Model}

A phenomenologically interesting intersecting D-brane model has been studied in Refs.~\cite{Chen:2007px,Chen:2007zu}. 
A variation of this model with a different hidden sector was also studied in Refs.~\cite{Maxin:2011ne,Chen:2007ms}. 
Type IIA orientifold string compactifications with intersecting 
D-branes (and their Type IIB duals with magnetized D-branes) 
have provided exciting geometric tools with which the MSSM may
be engineered.  While this approach may not allow a
first-principles understanding of why the SM gauge groups and
associated matter content arises, it may allow a deeper insight into
how the finer phenomenological details of the SM may emerge.
In short, D6-branes in Type IIA fill
(3+1)-dimensional Minkowski spacetime and wrap 3-cycles in the
compactified manifold, such that a stack of $N$ branes generates a
gauge group U($N$) [or U($N/2$) in the case of $T^6/(\Z_2 \times
\Z_2)$] in its world volume.  

In general, the 3-cycles wrapped by the stacks of D6-branes intersect
multiple times in the internal space, resulting
in a chiral fermion in the bifundamental representation localized at
the intersection between different stacks $a$ and $b$.  The multiplicity of such
fermions is then given by the number of times the 3-cycles intersect.
Each stack of D6-branes $a$ may 
intersect the orientifold images of other stacks $b'$, also resulting in fermions in
bifundamental representations.  Each stack may also intersect its own
image $a'$, resulting in chiral fermions in the symmetric and
antisymmetric representations.    
Non-chiral matter may also be present between stacks of D-branes which
do not intersect on one two-torus.  A zero
intersection number between two stacks of branes implies that the branes are
parallel on at least one torus.  At such kind of intersection
additional non-chiral (vector-like) multiplet pairs from $ab+ba$,
$ab'+b'a$, and $aa'+a'a$ can arise.  
Global consistency of the model requires certain constraints to be satisfied,
namely, Ramond-Ramond (R-R) tadpole cancellation and the preservation 
of $\mathcal{N}=1$ supersymmetry. In particular, the conditions for preserving
$\mathcal{N}=1$ supersymmetry fixes the complex structure parameters.

\begin{table}[tf]
\footnotesize
\renewcommand{\arraystretch}{1.0}
\caption{D6-brane configurations and intersection numbers for
a three-family Pati-Salam model on a Type-IIA $T^6 / (\Z_2
\times \Z_2)$ orientifold, with a tilted third two-torus.  The complete gauge
symmetry is $[{\rm U}(4)_C \times {\rm U}(2)_L \times {\rm
U}(2)_R]_{\rm observable} \times [ {\rm U}(2) \times {\rm USp}(2)^2]_{\rm hidden}$ and
$\mathcal{N}=1$ supersymmetry is preserved for $\chi_1=3$, $\chi_2=1$, $\chi_3=2$.}
\label{MI-Numbers}
\begin{center}
\begin{tabular}{|c||c|c|c|c|c|c|c|c|c|c|c|c|}
\hline
& \multicolumn{12}{c|}{${\rm U}(4)_C\times {\rm U}(2)_L\times {\rm
U}(2)_R \times {\rm U}(2) \times {\rm USp}(2)^2$}\\
\hline \hline  & $N$ & $(n^1,l^1)\times (n^2,l^2)\times

(n^3,l^3)$ & $n_{S}$& $n_{A}$ & $b$ & $b'$ & $c$ & $c'$& $d$ & $d'$ & 3 & 4 \\

\hline

     $a$&  8& $(0,-1)\times (1,1)\times (1,1)$ & \ 0 & \ 0  & 3 & 0 &
$-3$ & 0  & 0(2) & 0(1) & 0 & \ 0 \\

    $b$&  4& $(3,1)\times (1,~0)\times (1,-1)$ & \ 2 & $-2$  & - & - &
\ 0(6) & 0(1) & 1 & 0(1) & \ 0 & $-3$ \\

    $c$&  4& $(3,-1)\times (0,1)\times (1,-1)$ & $-2$ & \ 2  & - & - &
- & - & -1  & 0(1) &3 & \ 0 \\

    d&   4& $(1,0)\times (1,-1)\times (1,1)$ & 0 & 0 & - & - & - & - & - & - & -1 & 1
\\
\hline
    3&   2& $(0,-1)\times (~1,~0)\times (~0,~2)$ & \multicolumn{10}{c|}
{$\chi_1=3$}\\

    4&   2& $(0,-1)\times (~0,~1)\times (~2,0)$& \multicolumn{10}{c|}
{$\chi_2=1,~\chi_3=2$}\\

\hline

\end{tabular}

\end{center}

\end{table}

The set of D6 branes wrapping the cycles on a $T^6/(\Z_2 \times \Z_2)$
orientifold shown in Table~\ref{MI-Numbers} results in a
three-generation Pati-Salam model with additional hidden sectors.
The
full gauge symmetry of the model is given by $[{\rm U}(4)_C \times {\rm U}(2)_L \times {\rm
U}(2)_R]_{\rm observable} \times [ {\rm U}(2) \times {\rm USp}(2)^2]_{\rm hidden}$.  
As discussed in detail
in~\cite{Chen:2007px,Chen:2007zu}, with this configuration of D6 branes all R-R
tadpoles are canceled, K-theory constraints are satisfied, and
$\mathcal{N}=1$ supersymmetry is preserved.  Furthermore, the
tree-level MSSM gauge couplings are unified at the string scale.
Finally, the Yukawa matrices for quarks and leptons are rank 3 and it
is possible to obtain correct mass hierarchies and mixings.

Since ${\rm U}(N) = {\rm SU}(N)\times {\rm U}(1)$, associated with
each the stacks $a$, $b$, $c$, and $d$ are ${\rm U(1)}$ gauge groups,
denoted as ${\rm U}(1)_a$, ${\rm U}(1)_b$, ${\rm U(1)}_c$, and ${\rm U(1)}_d$.  
In general, these U(1)s are anomalous.  The anomalies associated with
these U(1)s are canceled by a generalized Green-Schwarz (G-S)
mechanism that involves untwisted R-R forms.  
As a result, the gauge bosons of these Abelian groups generically
become massive.  However, these U(1)s remain as global symmetries to all
orders in perturbation theory.  
These global U(1) symmetries
may also result in the forbidding of certain superpotential operators, such as Yukawa couplings
and those which mediate baryon and lepton number violation.
However, these {\it global} symmetries may be broken by nonperturbative effects, 
such as from D-brane instantons. 

Some linear combinations of $U(1)$s may also remain massless if certain conditions are satisfied.  
For the present model, precisely one linear combination 
has a massless gauge boson and is anomaly-
free:
\begin{equation}
{\rm U}(1)_X = {\rm U}(1)_a + 2\left[{\rm U}(1)_b + {\rm U}(1)_c + 3{\rm U}(1)_d
\right].
\end{equation}
Thus, the effective gauge symmetry of the model at the string scale is given by
\begin{equation}
{\rm SU}(4)_C \times {\rm SU}(2)_L \times {\rm SU}(2)_R \times {\rm
U}(1)_X \times \left[{\rm SU}(2) \times {\rm USp}(2)^2\right].
\end{equation}

\begin{figure}
	\centering
		\includegraphics[width=1.0\textwidth]{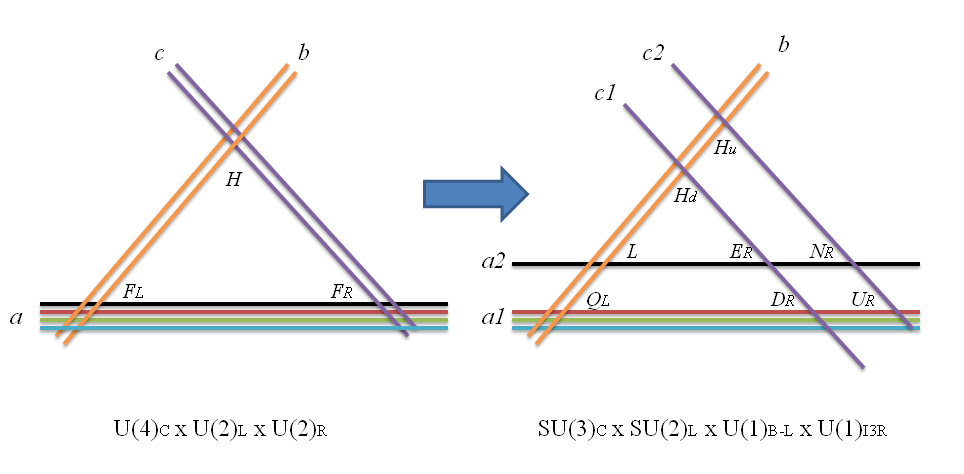}
		\caption{Breaking of the effective gauge symmetry via D-brane splitting. This process
		corresponds to assigning VEVs to adjoint scalars, which arise as open-string moduli
		associated with the positions of stacks {\it a} and {\it c} in the internal space.}
	\label{fig:Dsplitting}
\end{figure}

\begin{table}
[htb] \footnotesize
\renewcommand{\arraystretch}{1.0}
\caption{The chiral superfields, their multiplicities
and quantum numbers under the gauge symmetry $[{\rm SU}(3)_C \times {\rm SU}(2)_L \times {\rm
U}(1)_Y ]_{\rm observable} \times [ {\rm U}(1)_{I3V} \times {\rm USp}(2)^2]_{\rm hidden}$.
}
\label{MSSMSpectrum}
\begin{center}
\begin{tabular}{|c||c|c||c|c|c|c|c|c|c|c|c||c|}\hline
& Mult. & Quantum Number  & $Q_{I3R}$ & $Q_{B-L}$ & $Q_{3B+L}$ & $Q_{Y}$ & $Q_B$ & $Q_L$ & $Q_{Y'}$ & Field
\\
\hline\hline
$a1b$ & 3 & $(3,\overline{2},1,1,1,1)$ & 0 & 1/3 & 1 & 1/6 & 1/3 & 0 & 1/6 & $Q_L$\\
$a1c2$ & 3 & $(\overline{3},1,1,1,1,1)$ & -1/2 & -1/3 & -1  & -2/3 & -1/3 & 0 & -2/3 & $U_R$\\
$a1c1$ & 3 & $(\overline{3},1,1,1,1,1)$ & 1/2 & -1/3 & -1  & 1/3 & -1/3 & 0 & 1/3 & $D_R$\\
$a2b$ & 3 & $(1,\overline{2},1,1,1,1)$ & 0 & -1 & 1 & -1/2 &  0 & 1 & -1/2 & $L$\\
$a2c1$ & 3 & $(1,2,1,1,1,1)$ & 1/2 & 1 & -1  & 1 & 0 & -1 & 1 & $E_R$\\
$a2c2$ & 3 & $(1,2,1,1,1,1)$ & -1/2 & 1 & -1  & 0 & 0 & -1 & 0 & $N_R$\\
\hline
\end{tabular}
\end{center}
\end{table}

\begin{table}
[htb] \footnotesize
\renewcommand{\arraystretch}{1.0}
\caption{The chiral hidden sector superfields, their multiplicities
and quantum numbers under the gauge symmetry $[{\rm SU}(3)_C \times {\rm SU}(2)_L \times {\rm
U}(1)_Y \times ]_{\rm observable} \times [ {\rm U}(1)_{I3V} \times {\rm USp}(2)^2]_{\rm hidden}$.}
\label{HiddenSpectrum}
\begin{center}
\begin{tabular}{|c||c|c||c|c|c|c|c|c|c|c|c|c||c|}\hline
& Mult. & Quantum Number  & $Q_{I3R}$ & $Q_{B-L}$ & $Q_{3B+L}$ & $Q_{Y}$ & $Q_{I3V}$ & $Q_B$ & $Q_L$ & $Q_{Y'}$ & Field
\\
\hline\hline
$bd1$ & 1 & $(1,\overline{2},1,1,1,1)$ & 0 & 0 & -4  & 0 & 1/2 & -1 & -1 & 1/2 & $X_{bd1}$\\
$bd2$ & 1 & $(1,\overline{2},1,1,1,1)$ & 0 & 0 & -4  & 0 & -1/2 & -1 & -1 & -1/2 & $X_{bd2}$\\
$c1d1$ & 1 & $(1,1,1,1,1,1)$ & 1/2 & 0 & 4 & 1/2 & 1/2 & 1 & $1$ & 1 & $X_{c1d1}$\\
$c1d2$ & 1 & $(1,1,1,1,1,1)$ & 1/2 & 0 & 4 & 1/2 & -1/2 & 1 & $1$ & 0 & $X_{c1d2}$\\
$c2d1$ & 1 & $(1,1,1,1,1,1)$ & -1/2 & 0 & 4 & -1/2 & 1/2 & 1 & $1$ & 0 &  $X_{c2d1}$\\
$c2d2$ & 1 & $(1,1,1,1,1,1)$ & -1/2 & 0 & 4 & -1/2 & -1/2 & 1 & $1$ & -1 & $X_{c2d2}$\\
$b4$ & 3 & $(1,\overline{2},1,1,1,2)$ & 0 & 0 & -2 & 0 & 0 & -1/2 & -1/2 & 0 &$X_{b4}^i$ \\
$c13$ & 3 & $(1,1,\overline{2},1,2,1)$ & 1/2 & 0 & -2  &  1/2 & 0 & -1/2 & -1/2 &  1/2 & $X_{c13}^i$\\
$c23$ & 3 & $(1,1,\overline{2},1,2,1)$ & -1/2 & 0 & -2  &  -1/2 & 0 & -1/2 & -1/2 &  -1/2 & $X_{c23}^i$\\
$d13$ & 1 & $(1,1,1,1,2,1)$ & 0 & 0 & -6  & 0 & 1/2 & -3/2 & -3/2 & 1/2 & $X_{d13}$\\
$d23$ & 1 & $(1,1,1,1,2,1)$ & 0 & 0 & -6  & 0 & -1/2 & -3/2 & -3/2 & -1/2 & $X_{d23}$\\
$d14$ & 1 & $(1,1,1,1,\overline{2})$ & 0 & 0 & -6  & 0 & 1/2  & -3/2 & -3/2  & 1/2 & $X_{d14}$\\
$d24$ & 1 & $(1,1,1,1,\overline{2})$ & 0 & 0 & -6  & 0 & -1/2  & -3/2 & -3/2 & -1/2 & $X_{d24}$\\
$b_{S}$ & 2 & $(1,3,1,1,1,1)$ & 0 & 0 & -4  & 0 & 0 & -1 & -1 &  0 & $T_L^i$ \\
$b_{A}$ & 2 & $(1,1,1,1,1,1)$ & 0 & 0 & 4  & 0 & 0 & 1 & 1 & 0 & $S_L^i$\\
$c_{S}$ & 2 & $(1,1,1,1,1,1)$ & 0 & 0 & -4 & 0 & 0 & 0 & -1 & 0 &$T_R^i$\\
\hline
\end{tabular}
\end{center}
\end{table}

\begin{table}
[t] \footnotesize
\renewcommand{\arraystretch}{1.0}
\caption{The vector-like superfields not charged under $U(1)_{I3V}$, their multiplicities
and quantum numbers under the gauge symmetry 
$\left[SU(3)_C \times SU(2)_L \times U(1)_Y \right]_{observable} \left[U(1)_{I3V}\times USp(2)^2\right]_{hidden}$
and their charges under different $U(1)$ groups. Here $a$, $b$, $c$, etc. refer to different stacks of D-branes.}
\label{vector-likeSpectrumMSSM1}
\begin{center}
\begin{tabular}{|c||c|c||c|c|c|c|c|c|c|c|c|c||c|}\hline
& Mult. & Quantum Number &  $Q_{I3R}$ & $Q_{B-L}$ & $Q_{3B+L}$ & $Q_{Y}$ & $Q_B$ & $Q_L$ & $Q_{Y'}$ & Field
\\
\hline\hline
$a1b'$ & 3 & $(3,2,1,1,1,1)$ & \ 0 & \ 1/3 & -3 & 1/6 & \ -2/3 & -1 & 1/6 & $XQ^i_L$ \\
& 3 & $(\overline{3},\overline{2},1,1,1,1)$ & $0$ & -1/3 & 3 & -1/6 & 2/3 & 1 & -1/6 &
$\overline{XQ}^i_L$ \\
\hline
$a2b'$ & 3 & $(1,2,1,1,1,1)$ & \ 0 & \ -1 & -3 & -1/2 & \ -1 & 0 & -1/2 & $XL^i_L$ \\
& 3 & $(1,\overline{2},1,1,1,1)$ & $0$ & 1 & 3 & 1/2 & 1 & 0 & 1/2 & 
$\overline{XL}^i_L$ \\
\hline
$a1c1'$ & 3 & $(3,1,1,1,1,1)$ & \ 1/2 & 1/3 & \ -3  & 2/3 &  \ -2/3 & -1 & 2/3 & $XU^i_R$ \\
& 3 & $(\overline{3},1,1,1,1,1)$ & -1/2 & -1/3 & 3 & -2/3 & 2/3 & 1 & -2/3
& $\overline{XU}^i_R$\\
\hline
$a1c2'$ & 3 & $(3,1,1,1,1,1)$ & \ -1/2 & 1/3 & \ -3  & -1/3 &  \ -2/3 & -1 & -1/3 & $XD^i_R$ \\
& 3 & $(\overline{3}, 1,1,1,1,1)$ & 1/2 & -1/3 & 3 & 1/3 & 2/3 & 1 & 1/3 
& $\overline{XD}^i_R$\\
\hline
$a2c1'$ & 3 & $(1,1,1,1,1,1)$ & \ 1/2 & -1 & \ -3  & 0 &  \ -1 &   0 & 0 & $XN^i_R$ \\
& 3 & $(1,1,1,1,1,1)$ & -1/2 & 1 & 3 & 0 & 1 & 0 & 0 
& $\overline{XN}^i_R$\\
\hline
$a2c2'$ & 3 & $(1,1,1,1,1,1)$ & \ -1/2 & -1 & \ -3  & -1 &  \ -1 &  0 & -1 & $XE^i_R$ \\
& 3 & $(1,1,1,1,1,1)$ & 1/2 & 1 & 3 & 1 & 1 & 0 & 1 
& $\overline{XE}^i_R$\\
\hline
$bc1$ & 6 & $(1,2,1,1,1,1)$ & -1/2 & 0 & 0  & -1/2 & 0 &  0 & -1/2 & $H_d^i$\\
      & 6 & $(1,\overline{2},1,1,1,1)$ & 1/2 & 0 & 0  & 1/2 & 0 &  0 & 1/2 & $\overline{H}_d^i$\\
\hline			
$bc2$ & 6 & $(1,2,1,1,1,1)$ & 1/2 & 0 & 0  & 1/2 & 0 & 0 &  1/2 & $H_u^i$\\
      & 6 & $(1,\overline{2},1,1,1,1)$ & -1/2 & 0 & 0  & -1/2 & 0 & 0 &  -1/2 & $\overline{H}_u^i$\\
\hline
$bc1'$ & 1 & $(1,\overline{2},1,1,1,1)$ & -1/2 & 0 & 4  & -1/2 & 1 &  1 & -1/2 & $\overline{\mathcal{H}}_1$\\ 
       & 1 & $(1,2,1,1,1,1)$ & 1/2 & 0 & -4  & 1/2 & -1 &  -1 & 1/2 & $\mathcal{H}_1$\\
\hline			
$bc2'$ & 1 & $(1,\overline{2},1,1,1,1)$ &  1/2 & 0 & 4  &  1/2 & 1 &  1 &  1/2 & $\overline{\mathcal{H}}_2$\\
       & 1 & $(1,2,1,1,1,1)$ &  -1/2 & 0 & -4  &  -1/2 & -1 &  -1 &  -1/2 & $\mathcal{H}_2$\\
\hline			
\end{tabular}
\end{center}
\end{table}

The gauge symmetry is first broken by splitting the D-branes as $a\rightarrow a1 + a2$ with $N_{a1}=6$ and $N_{a2}=2$, and $c\rightarrow c1 + c2$ with $N_{c1}=2$ and $N_{c2}=2$, and $d\rightarrow d1 + d2$ with$ N_{d1}=2$ and $N_{d2}=2$, as shown schematically in Fig.\ref{fig:Dsplitting}. 
After splitting the D6-branes, the gauge symmetry of the observable sector is 
\begin{equation}
SU(3)_C \times SU(2)_L \times U(1)_{I3R} \times U(1)_{B-L} \times U(1)_{3B+L} \times U(1)_{I3V},
\end{equation}
where 
\begin{eqnarray}
U(1)_{I3R} &=& \frac{1}{2}(U(1)_{c1} - U(1)_{c2}), \ \ \  
U(1)_{B-L} = \frac{1}{3}(U(1)_{a1} - 3U(1)_{a2}), \\ \nonumber
U(1)_{I3V} &=& \frac{1}{2}(U(1)_{d1} - U(1)_{d2}),
\end{eqnarray}
and
\begin{equation}
U(1)_{3B+L}= -[U(1)_{a1}+U(1)_{a2} + 2(U(1)_{b}+U(1)_{c1}+U(1)_{c2}+3U(1)_{d1}+3U(2)_{d2})],
\end{equation}

The gauge symmetry must be further broken to the SM, with the possibility of one or more additional U(1) gauge symmetries. In particular, 
the $U(1)_{B-L}\times U(1)_{I_{3R}}\times U(1)_{3B+L}$ gauge symmetry may be broken by assigning VEVs to the right-handed neutrino fields $N_R^i$.  In this case, the gauge symmetry is broken to
\begin{equation}
\left[SU(3)_C \times SU(2)_L \times U(1)_Y \times U(1)_B  \right]_{observable} \times \left[U(1)_{I3V} \times USp(2)^2\right]_{hidden}
\label{MSSM_B}
\end{equation}
where 
\begin{eqnarray}
&U(1)_Y = \frac{1}{6}\left[U(1)_{a1}-3U(1)_{a2}+3U(1)_{c1}-3U(1)_{c2}\right] \\ \nonumber
& = \frac{1}{2}U(1)_{B-L}+U(1)_{I3R},
\end{eqnarray}
and
\begin{eqnarray}
&U(1)_B = \frac{1}{4}[U(1)_{B-L} + U(1)_{3B+L}]\\ \nonumber
&= -[\frac{1}{6}U(1)_{a1}+\frac{1}{2}(U(1)_{a2} + U(1)_{b}+U(1)_{c1}+U(1)_{c2}+3U(1)_{d})].
\end{eqnarray}
Alternatively, the gauge symmetry may also be broken by assigning VEVs to the vector-like fields
$XN^i_R$ and $\overline{XN}^i_R$ in the $a_2c_1$ sector shown in Table~\ref{vector-likeSpectrumMSSM2}. 
The gauge symmetry in this case is then
\begin{equation}
\left[SU(3)_C \times SU(2)_L \times U(1)_Y \times U(1)_L \right]_{observable} \times \left[U(1)_{I3V} \times USp(2)^2\right]_{hidden}
\label{MSSM_L}
\end{equation}
where 
\begin{eqnarray}
&U(1)_L = \frac{1}{4}[-3U(1)_{B-L} + U(1)_{3B+L}].
\end{eqnarray}
From Table~\ref{MSSMSpectrum}, it may be seen that $U(1)_B$ and $U(1)_L$ count 
baryon number and lepton respectively for the chiral fields, although this is 
not the case for the vector-like fields.  
The $U(1)_Y \times U(1)_B \times U(1)_{I3V}$ and $U(1)_Y \times U(1)_L \times U(1)_{I3V}$
gauge symmetries may also be broken by assigning VEVs to some of the 
vector-like singlet fields $\varphi21_i$, $\varsigma21$, $\psi12$, and $\psi21$ shown in 
Table~\ref{vector-likeSpectrumMSSM3}.  
After this breaking, one anomaly-free linear combination
remains:
\begin{equation}
U(1)_{Y'} = \frac{1}{6}\left[U(1)_{a1}-3U(1)_{a2}+3U(1)_{c1}-3U(1)_{c2}+3U(1)_{d1}-3U(1)_{d2}\right].
\end{equation}
The VEVs assigned to the vector-like singlets may be string scale or 
just below the string scale, so that below the string scale
the gauge symmetry is 
\begin{equation}
\left[SU(3)_C \times SU(2)_L \times U(1)_{Y'} \right]_{observable} \times \left[USp(2)^2\right]_{hidden},
\label{MSSM}
\end{equation}
with $U(1)_{Y'}$ being identified with the SM hypercharge.  
We will further assume that all exotic matter, shown in Table~\ref{HiddenSpectrum}, may become massive, as shown in Ref.~\cite{Chen:2007ms}. The resulting low-energy field content is shown in Tables~\ref{MSSMSpectrum} and along with their charges under 
$U(1)_{I3R}$, $U(1)_{B-L}$, $U(1)_{3B+L}$, $U(1)_Y$, $U(1)_{I3V}$, $U(1)_B$, and $U(1)_L$.
It should be noted that there are several fields present which are SM singlets.  

Finally, it is possible to calculate the gauge couplings at the string scale.  For this model, it is found that the 
tree-level gauge couplings are unified at the string scale:
\begin{equation}
g^2_s = g^2_w = \frac{5}{3} g^2_Y = 2 g^2_{Y'},
\end{equation}
where the unification with $g^2_{Y'}$ is non-canonical~\cite{Chen:2007zu}.  
Moreover, the hidden sector gauge groups $USp(2)_3$ and $USp(2)_4$ will 
become strongly coupled near the string scale, thus decoupling matter
charged under these groups~\cite{Chen:2007zu}.  

\begin{table}
[t] \footnotesize
\renewcommand{\arraystretch}{1.0}
\caption{The vector-like quarks and $SU(2)_L$ doublets charged under $U(1)_{I3V}$, their multiplicities
and quantum numbers under the gauge symmetry 
$\left[SU(3)_C \times SU(2)_L \times U(1)_Y \right]_{observable} \left[ U(1)_{I3V} \times USp(2)^2\right]_{hidden}$
and their charges under different $U(1)$ groups. Here $a$, $b$, $c$, etc. refer to different stacks of D-branes.
}
\label{vector-likeSpectrumMSSM2}
\begin{center}
\begin{tabular}{|c||c|c||c|c|c|c|c|c|c|c|c|c|c||c|}\hline
& Mult. & Quantum Number &  $Q_{I3R}$ & $Q_{B-L}$ & $Q_{3B+L}$ & $Q_{Y}$ & $Q_{I_{3V}}$ & $Q_B$ & $Q_L$ & $Q_{Y'}$ & Field
\\
\hline
\hline
$a1d1$ & 2 & $(3,1,1,1,1,1)$ & \ 0 & 1/3 & 5 & 1/6 & 1/2  & 4/3 & 1 & 2/3 & $\varphi11_i$\\
& 2 & $(\overline{3},1,1,1,1,1)$ & 0 & -1/3 & -5 & -1/6 & -1/2 &  -4/3 &  -1  & -2/3 &  $\overline{\varphi11}_i$\\ 
\hline
$a1d2$ & 2 & $(3,1,1,1,1,1)$ & \ 0 & 1/3 & 5 & 1/6 &  -1/2 & 4/3 & 1 & -1/3 & $\varphi12_i$\\
& 2 & $(\overline{3},1,1,1,1,1)$ & 0 & -1/3 & -5 & -1/6 & 1/2 &  -4/3   &  -1 & 1/3 & $\overline{\varphi12}_i$\\ 
\hline
$a1d1'$ & 1 & $(3,1,1,1,1,1)$ & \ 0 & 1/3 & -7 & 1/6 & 1/2 & -5/3 & -2 & 2/3 & $\varsigma11$ \\
& 1 & $(\overline{3},1,1,1,1,1)$ & 0 & -1/3 & 7 & -1/6 & -1/2 & 5/3 & 2 & -2/3 & $\overline{\varsigma11}$\\
\hline
$a1d2'$ & 1 & $(3,1,1,1,1,1)$ & \ 0 & 1/3 & -7 & 1/6 & -1/2 & -5/3 & -2 & -1/3 & $\varsigma12$ \\
& 1 & $(\overline{3},1,1,1,1,1)$ & 0 & -1/3 & 7 & -1/6 & 1/2 & 5/3 & 2 &   1/3 & $\overline{\varsigma12}$\\
\hline	
$bd1'$ & 1 & $(1,2,1,1,1,1)$ & 0 & 0 & -8 & 0 & 1/2 & -2 & -2 & 1/2 & $\xi_1$ \\
     & 1  & $(1,\overline{2},1,1,1,1)$ & 0 & 0 & 8  & 0 & -1/2 & 2 & 2 & -1/2 &	$\overline{\xi}_1$\\
\hline
$bd2'$ & 1 & $(1,2,1,1,1,1)$ & 0 & 0 & -8 & 0 & -1/2 & -2 & -2 & -1/2 & $\xi_2$ \\
			 & 1  & $(1,\overline{2},1,1,1,1)$ & 0 & 0 & 8  & 0 & 1/2 & 2 & 2 & 1/2 &	$\overline{\xi}_2$\\
\hline			
\end{tabular}
\end{center}
\end{table}

\begin{table}
[t] \footnotesize
\renewcommand{\arraystretch}{1.0}
\caption{The vector-like singlets charged under $U(1)_{I3V}$,their multiplicities
and quantum numbers under the gauge symmetry 
$\left[SU(3)_C \times SU(2)_L \times U(1)_Y \right]_{observable} \left[U(1)_{I3V}\times USp(2)^2\right]_{hidden}$
and their charges under different $U(1)$ groups. Here $a$, $b$, $c$, etc. refer to different stacks of D-branes.}
\label{vector-likeSpectrumMSSM3}
\begin{center}
\begin{tabular}{|c||c|c||c|c|c|c|c|c|c|c|c|c||c|}\hline
& Mult. & Quantum Number &  $Q_{I3R}$ & $Q_{B-L}$ & $Q_{3B+L}$ & $Q_{Y}$ & $Q_{I_{3V}}$ & $Q_B$ & $Q_L$ & $Q_{Y'}$ & Field
\\
\hline
\hline
$a2d1$ & 2 & $(1,1,1,1,1,1)$ &  0 & -1 & 5 & -1/2  & 1/2 & 1 & 1/2 & 0 & $\varphi21_i$\\
& 2 & $(1,1,1,1,1,1)$ & 0 & 1 & -5 & 1/2 & -1/2 &  -1   & -1/2 &  0 & $\overline{\varphi21}_i$\\ 
\hline
$a2d2$ & 2 & $(1,1,1,1,1,1)$ &  0 & -1 & 5 & -1/2 &  -1/2 & 1 & 1/2 & -1 & $\varphi22_i$\\
& 2 & $(1,1,1,1,1,1)$ & 0 & 1 & -5 & 1/2 & 1/2 & -1   &  -1/2 & 1 & $\overline{\varphi22}_i$\\ 
\hline
$a2d1'$ & 1 & $(1,1,1,1,1,1)$ &  0 & -1 & -7 & -1/2 & 1/2 & -2 & -1 & 0 & $\varsigma21$ \\
& 1 & $(1,1,1,1,1,1)$ & 0 & 1 & 7 & 1/2 & -1/2 & 2 & 1  & 0 & $\overline{\varsigma21}$\\
\hline
$a2d2'$ & 1 & $(1,1,1,1,1,1)$ &  0 & -1 & -7 & -1/2 & -1/2 & -2 & -1 & -1 & $\varsigma22$ \\
& 1 & $(1,1,1,1,1,1)$ & 0 & 1 & 7 & 1/2 & 1/2 & 2 & 1 & 1 & $\overline{\varsigma22}$\\
\hline
$c1d1'$ & 1 & $(1,1,1,1,1,1)$ & 1/2 & 0 & -8 & 1/2 & 1/2 & -2 & -2 & 1 & $\psi11$ \\
     & 1  & $(1,1,1,1,1,1)$ & -1/2 & 0 & 8  & -1/2 & -1/2 & 2 & 2 & -1 & $\overline{\psi11}$\\	
\hline		
$c1d2'$ & 1 & $(1,1,1,1,1,1)$ & 1/2 & 0 & -8 & 1/2 & -1/2 & -2 & -2 & 0 & $\psi12$ \\
     & 1  & $(1,1,1,1,1,1)$ & -1/2 & 0 & 8  & -1/2 & 1/2 & 2 & 2 & 0 & $\overline{\psi12}$\\	
\hline		
$c2d1'$ & 1 & $(1,1,1,1,1,1)$ & -1/2 & 0 & -8 & -1/2 & 1/2 & -2 & -2 & 0 & $\psi21$ \\
     & 1  & $(1,1,1,1,1,1)$ & 1/2 & 0 & 8  & -1/2 & -1/2 & 2 & 2 & 0 & $\overline{\psi21}$\\	
\hline		
$c2d2'$ & 1 & $(1,1,1,1,1,1)$ & -1/2 & 0 & -8 & -1/2 & -1/2 & -2 & -2 & -1 & $\psi22$ \\
     & 1  & $(1,1,1,1,1,1)$ & 1/2 & 0 & 8  & 1/2 & 1/2 & 2 & 2 & 1 & $\overline{\psi22}$\\	
\hline		
\end{tabular}
\end{center}
\end{table}

\section{Gauge Coupling Unification and Vector-like Matter}

Vector-like matter appears in intersecting/magnetized D-brane models on toroidal orientifolds between stacks of D-branes which 
do not intersect.  The mass of such vector-like states depends upon the separation between the stacks of D-branes in the 
internal space.  As such, it is generically massive.  Only stacks of D-branes which are directly on top of one another have 
massless vector-like states between them.  In the model studied in Section II, the toroidal orientifold consist of a six-torus which
is factorizable, $\mathbf{T}^6 = \mathbf{T}^2 \times \mathbf{T}^2 \times \mathbf{T}^2$.  If two stacks of D-branes are parallel 
on one two-torus, then vector-like matter appears in the bifundamental representation of the gauge groups within the world-volume
of each stack. If the two stacks are not separated on the two torus on which they are parallel, the vector-like multiplets
are massless.  However, these states become massive if the stacks are separated.   

The most straightforward way to obtain the diphoton excesses is with a SM singlet scalar with a mass $\sim~750$~GeV coupled to vector-like quarks.  The coupling
to vector-like quarks is necessary in order to produce the scalar via gluon fusion and to allow the decay of the scalar into diphotons.  
The requisite vector-like quarks are indeed present in the model.  In particular, 
the vector-like quarks and leptons in the $ab'$ and  
$ac'$ sector fill $\mathbf{16}$ and $\overline{\mathbf{16}}$ spinorial representations of SO(10), 
or equivalently $\mathbf{5}+\overline{\mathbf{5}}+\mathbf{10}+\overline{\mathbf{10}}+\mathbf{1} + \overline{\mathbf{1}}$ of SU(5).  
For example,
\begin{eqnarray}
\label{abvector}
\mathbf{5}^i + \overline{\mathbf{5}}^i &=& \left\{(XD^i_R, \overline{XD}^i_R), (XL^i_L, \overline{XL}^i_L)\right\}, \\ \nonumber
\mathbf{10}^i + \overline{\mathbf{10}}^i &=& \left\{(XQ^i_L, \overline{XQ}^i_L), (XU^i_R, \overline{XU}^i_R), (XE^i_R, \overline{XE}^i_R)\right\}, \\ \nonumber
\mathbf{1}^i + \overline{\mathbf{1}}^i   &=& \left\{XN^i_R, \overline{XN}^i_R\right\}.
\end{eqnarray}

It is well-known that gauge coupling unification may be preserved at the 1-loop level if the extra matter comes in complete 
representations of $SU(5)$.  However, at the 2 or 3-loop level, a Landau pole may appear.  This restricts the number of $SU(5)$ 
multiplets which may remain light to either one $(\mathbf{10}+\mathbf{\overline{10}})$ or
three copies of $(\mathbf{5}+\mathbf{\overline{5}})$.    
In addition, any number of SM singlets may be present in the light spectrum.  
Generically, there are many more vector-like states in the spectrum, as can be seen
from Tables~\ref{vector-likeSpectrumMSSM1},\ref{vector-likeSpectrumMSSM2}, and 
\ref{vector-likeSpectrumMSSM3}. Thus, in order to preserve gauge coupling unification,
many of these states must obtain string-scale masses, while simultaneously maintaining
light masses for either of the two cases stated above. 
Recalling that vector-like matter appears
between stacks of D-branes which are parallel on one two-torus, the masses of these vector-like states depend on 
the separation between 
the stacks on the two-torus on which they are parallel. 
So, it is possible to give string-scale masses to some of the vector-like states by
separating the two stacks of D-branes on the two-torus on which they are parallel and
where these vector-like states are localized.  

Is it possible to choose the position of the stacks of D-branes so that
only the $\mathbf{5}+\overline{\mathbf{5}}$ or $\mathbf{10}+\overline{\mathbf{10}}$ fields remain light?  The answer to this question
is yes.  As an example, let us consider just the fields in the $ab'$ and $ac'$ sectors, where
it should be noted that these stacks of D-branes are parallel on the third 
two-torus.  Thus, for example the fields $(XQ^i_L, \overline{XQ}^i_L)$ may become massive by separating stacks $a1$ and $b'$ on 
the third torus.  Similarly, the fields $(XU^i_R, \overline{XU}^i_R)$ may become massive by separating stacks $a1$ and $c1'$, while
$(XE^i_R, \overline{XE}^i_R)$ become massive if stacks $a2$ and $c2'$ are separated on the third two-torus.  The fields 
$(XD^i_R, \overline{XD}^i_R)$ may remain massless if stacks $a1$ and $c2'$ overlap on the third two-torus, and the same is 
true for $(XL^i_L, \overline{XL}^i_L)$ if stacks $a2$ and $b'$ overlap on the third two-torus.  
Thus with this configuration, only the fields in the $\mathbf{5}^i + \overline{\mathbf{5}}^i$  and $\mathbf{1}^i + \overline{\mathbf{1}}^i$ 
(with $i=1\ldots3$)
representations of Eq.~\ref{abvector} remain light.  

In addition, there are additional vector-like quarks and leptons in the 
spectrum in the $ad$ and $ad'$ sectors, as shown in 
Tables~\ref{vector-likeSpectrumMSSM2} and~\ref{vector-likeSpectrumMSSM3}. These fields may also become massive by separating
the relevant stacks of D-branes on the two-torus where they are parallel. For example, stacks $a1$ and $a2$ are parallel with 
stacks $d1$ and $d2$ on the third two-torus, and so the vector-like matter in these sectors may be eliminated if these stacks are
separated on the third two-torus. Similarly, stacks $a1$ and $a2$ are parallel with stacks $d1'$ and $d2'$ on the second two-torus,
and the vector-like matter in this sector may be eliminated by displacing these stacks on the second two-torus.

On the other hand, it must also be kept in mind that splitting the stacks in this fashion may 
affect the Higgs sectors, which  are also vector-like.  For example 
the Higgs fields $H_u$ and $H_d$ arise from vector-like matter in the
$bc1$ and $bc2$ sectors respectively.  Stackss $a1$ and $c2'$ 
must overlap to keep $(XE_R^i, \overline{XE}_R^i)$ light while 
stacks $a1$ and $b'$ must be separated to give masses to the fields 
$(XQ_L^i, \overline{XQ}_L^i$) such that the light vector-like fields may be
placed in $\mathbf{5}+\overline{\mathbf{5}}$ representations. 
However, this implies that stacks $b$ and $c2$ 
must be separated and thus that the Higgs field $H_d$ is not 
present in the light spectrum. Clealy, then it is not 
possible to eliminate the fields in the $\mathbf{10}+\overline{10}$
while keeping the fields in the $\mathbf{5}+\overline{5}$ 
without also eliminating some of the Higgs field.  
Note that this constraint only applies to those fields in the 
$ab'$ and $ac'$ sectors.  In the next section, we shall also
consider the vector-like matter in the $ad$ and $ad'$ sectors.
We shall find that it is possible to have just vector-like matter 
in the $\mathbf{5}+\overline{5}$ representation 
without eliminating some of the Higgs fields as a by-product.

\section{MSSM Singlets Coupled to Vector-like Matter}

We have seen in the previous section that the model contains vector-like quarks and leptons
in the $ab'$ and $ac'$ sectors which may be grouped into complete representations of $SU(5)$.  
Furthermore, by displacing
the stacks of D-branes, it is possible to eliminate some of these fields from the spectrum
so that only three copies of $\mathbf{5}+\mathbf{\overline{5}}$ remain light.
In this way, gauge coupling unification may be maintained while also avoiding the Landau pole problem.  
However, upon inspection, none of these vector-like quarks and leptons appear to have 
Yukawa couplings with any of the singlet fields present in the model.  Therefore, they may not be involved
in producing the diphoton excesses.  
However, this is 
not the case if we examine the vector-like quarks and leptons present in the 
$ad$ and $ad'$ sectors.   

Let us turn our attention to the SM singlet fields 
$X_{c1d2}$ and $X_{c2d1}$ shown in Table~\ref{HiddenSpectrum}.  These fields have Yukawa couplings
with the vector-like quarks in the $ad$ and $ad'$ sectors:
\begin{eqnarray}
W_3 &=& c_1^i  \cdot \overline{\psi12}\cdot X_{c1d2} \cdot \varphi12^i \cdot\varsigma12 +  c_2^i \cdot \overline{\psi21}\cdot X_{c2d1} \cdot \varphi11^i \cdot \varsigma11, 
\label{VectorQuarkSingletYukawa}	
\end{eqnarray}
where $\left\langle \overline{\psi12}\right\rangle$ and $\left\langle \overline{\psi21}\right\rangle$ are near the string scale $M_{St}$
as discussed in Section I.  In addition, there are Yuakwa couplings between these singlets and some of the vector-like doublets:
\begin{eqnarray}
W_2 &=& k_1 X_{c1d2} \overline{\mathcal{H}}_1 \xi_2 + k_2 X_{c2d1} \overline{H}_2 \xi_1. 
\end{eqnarray}
Finally, there are additional Yukawa couplings between $X_{c1d2}$ and $X_{c2d1}$ and the singlet fields in the 
$a2d$ and $a2d'$ sectors:
\begin{eqnarray}
W_1 &=& d_1^i  \cdot \overline{\psi12}\cdot X_{c1d2} \cdot \varphi22^i \cdot\varsigma22 +  d_2^i \cdot \overline{\psi21}\cdot X_{c2d1} \cdot \varphi21^i \cdot \varsigma21. 
\label{VectorSingletSingletYukawa}	
\end{eqnarray}

Thus, the required couplings of the SM singlet fields to vector-like quarks are present  
so that the singlet fields may be produced by gluon fusion and decay to diphotons. In addition, the singlet fields
have additional couplings to other doublet and singlet fields.  
Note that the vector-like 
quarks involved in these couplings combined with additional vector-like matter may be placed in complete representations of $SU(5)$
by replacing the right-handed 
vector-like quarks, leptons, and singlets in Eq.~\ref{abvector} with some of the quarks, leptons and singlets of
Tables~\ref{vector-likeSpectrumMSSM2} and~\ref{vector-likeSpectrumMSSM3}.  For example, making the interchanges
\begin{eqnarray}
(XD^i_R, \overline{XD}^i_R) \rightarrow (\varphi12^i, \overline{\varphi12}^i), \\ \nonumber
(XU^i_R, \overline{XU}^i_R) \rightarrow (\varphi11^i, \overline{\varphi11}^i), \\ \nonumber
(XE^i_R, \overline{XE}^i_R) \rightarrow (\varphi22^i, \overline{\varphi22}^i), \\ \nonumber
(XN^i_R, \overline{XN}^i_R) \rightarrow (\varphi21^i, \overline{\varphi21}^i), \\ \nonumber
(XL^i_L, \overline{XL}^i_L) \rightarrow (\mathcal{H}^i, \overline{\mathcal{H}}^i), \\ \nonumber
\label{advector0}
\end{eqnarray}
we have 
\begin{eqnarray}
\label{advector1}
\mathbf{5}^i + \overline{5}^i &=& \left\{(\varphi12_i, \overline{\varphi12}^i), (\mathcal{H}^i, \overline{\mathcal{H}}^i)\right\}, \\ \nonumber
\mathbf{10}^i + \overline{10}^i &=& \left\{(XQ^i_L, \overline{XQ}^i_L), (\varphi11_i^i, \overline{\varphi11}^i), (\varphi22^i, \overline{\varphi22}^i)\right\}, \\ \nonumber
\mathbf{1}^i + \overline{\mathbf{1}}^i   &=& \left\{\varphi21^i, \overline{\varphi21}^i\right\}.
\end{eqnarray}
where $i=1\ldots 2$, and making the interchanges
\begin{eqnarray}
(XD^3_R, \overline{XD}^3_R) \rightarrow (\varsigma12, \overline{\varsigma12}), \\ \nonumber
(XU^3_R, \overline{XU}^3_R) \rightarrow (\varsigma11, \overline{\varsigma11}),  \\ \nonumber
(XE^3_R, \overline{XE}^3_R) \rightarrow (\varsigma22, \overline{\varsigma22}),  \\ \nonumber 
(XN^3_R, \overline{XN}^3_R) \rightarrow (\varsigma21, \overline{\varsigma21}),  \\ \nonumber
(XL^3_L, \overline{XL}^3_R) \rightarrow (\xi_1, \overline{\xi_1}),  \\ \nonumber
\label{advector15}
\end{eqnarray}
we have
\begin{eqnarray}
\label{advector2}
\mathbf{5} + \overline{\mathbf{5}} &=& \left\{(\varsigma12, \overline{\varsigma12}), (\xi_1, \overline{\xi_1})\right\}, \\ \nonumber
\mathbf{10} + \overline{\mathbf{10}} &=& \left\{(XQ^3_L, \overline{XQ}^3_L), (\varsigma11, \overline{\varsigma11}), (\varsigma22, \overline{\varsigma22})\right\}, \\ \nonumber
\mathbf{1} + \overline{\mathbf{1}}   &=& \left\{\varsigma21, \overline{\varsigma21}\right\}.
\end{eqnarray}

In order  
to have just three $\mathbf{5}+\mathbf{\overline{5}}$ multiplets + additional singlets in the light spectrum,
the fields grouped into the $\mathbf{10}+\mathbf{\overline{10}}$ multiplets must become
massive, as well as any additional vector-like states.   
To eliminate the vectorlike fields in the $ab'$ and $ac'$ sectors from the light spectrum,
we must require that stacks $a1$ and $a2$ are separated from stacks $b'$, $c1'$, and $c2'$
on the third two-torus. Futhermore, to eliminate the fields in the above $\mathbf{10}+\overline{\mathbf{10}}$,
we require that stack $a1$ be separated from stack $d1$ 
on the third two-torus and from $d1'$ on the second two-torus.  Let us also require 
that stack $a2$ be separated from both stacks $d1$, $d1'$, $d2$ and $d2'$. 

To keep the fields in the above $\mathbf{5}+\mathbf{\overline{5}}$ representations light,
we require that stack $a1$ overlap with stack $d2$ on the third two-torus and with stack
$d2'$ on the second two torus.  In addition, we require that stack $b$ overlap stacks
$c1$, $c1'$, $c2$, and $c2'$ as well as stacks $d1'$.  This configuration assures
that the Higgs fields $H^i_u$ and $H^i_d$ are present in the spectrum.  
In addition, stack $b$ may not overlap 
stack $d1'$ since stack $a1$ overlaps stack $d1$.  Since stack $a1$ and $b'$ 
are separated, this implies that stacks $b$ and $d1'$ are also separated.

Then, the fields in the light spectrum consist of the following fields with quantum numbers
under the $SU(3)_C \times SU(2)_L \times U(1)_Y$ gauge symmetry shown: 
\begin{eqnarray}
\label{lightreps}
(XD_{1,2},XD^c_{1,2}) &\equiv&  2\times (\varphi12, \overline{\varphi12}) = 2\times\left\{ (\overline{\mathbf{3}},\mathbf{1},\mathbf{\frac{1}{3}}) + (\mathbf{3},\mathbf{1},\mathbf{-\frac{1}{3}})\right\}, \\ \nonumber 
(XD_{3},XD^c_{3}) &\equiv& 1\times(\varsigma12, \overline{\varsigma12}) = \left\{(\overline{\mathbf{3}},\mathbf{1},\mathbf{\frac{1}{3}}) + (\mathbf{3},\mathbf{1},\mathbf{-\frac{1}{3}})\right\}, \\ \nonumber
 (XL_{L_{1,2}}, XL^c_{L_{1,2}}) &\equiv& 2\times (\mathcal{H}_1, \overline{\mathcal{H}_1}) = \left\{(\mathbf{1},\overline{\mathbf{2}},\mathbf{-\frac{1}{2}}) + (\mathbf{1},\mathbf{2},\mathbf{\frac{1}{2}})\right\}, \\ \nonumber
(XL_{L_{3}}, XL^c_{L_{3}}) &\equiv& 1\times (\xi_1, \overline{\xi_1}) = \left\{(\mathbf{1},\overline{\mathbf{2}},\mathbf{-\frac{1}{2}}) + (\mathbf{1},\mathbf{2},\mathbf{\frac{1}{2}})\right\},
\end{eqnarray}
plus additional singlets.  As shown in Eq.~\ref{VectorQuarkSingletYukawa}, the vector-like quarks 
$(\varphi12^{1,2}, \overline{\varphi12}^{1,2})$ and $(\varsigma12, \overline{\varsigma12})$ have Yukawa 
couplings with the singlet field $S \equiv X_{c1d2}$.  Thus, this singlet field may be produced via loops
involving these vector-like quarks as well as decay to diphotons via gluon fusion.  In addition,
the singlet field S has couplings to the doublets $(XL_L, XL^c_L)$ to which it may also
decay.

Using the notation of Eq.~\ref{lightreps}, the superpotential for the extra vector-like states
and the singlet $S$ is
\begin{eqnarray}
  W&=&  \lambda_D S XD XD^c + \lambda_L S XL XL^c
   \nonumber \\
 && + M_{XD} XD XD^c + M_{XL} XL XL^c~.~
\end{eqnarray}
The corresponding supersymmetry breaking soft terms are
\begin{eqnarray}
  V_{\rm soft} &=& 
   {\widetilde M}^2_{XD} (|{\widetilde{XD}}|^2+ |{\widetilde{XD}^c}|^2)
  + {\widetilde M}^2_{XL} (|{\widetilde{XL}}|^2+ |{\widetilde{XL}^c}|^2)
  \nonumber \\
  &&   -\left(\lambda_D A_D S \widetilde{XD} {\widetilde{XD}^c}
  \right. \nonumber \\   && \left.
+\lambda_L A_L S \widetilde{XL} {\widetilde{XL}^c}
 \right. \nonumber \\   && \left.
+ B_{XD} M_{XD} XD XD^c + B_{XL} M_{XL} XL XL^c
+{\rm H.C.}\right)~.~\,
\end{eqnarray}

\section{The Diphoton Excesses}

\begin{table}[htbp]
\centering
\tiny
\caption{Decay widths and production cross-sections for a total decay width of $\Gamma = 5$~GeV for some sample points. All masses and decay widths are in GeV. The cross-sections are in femtobarns (fb). The ${\rm Br}_{\rm DM}$ represents the branching ratio allocated to dark matter. For simplicity, we assume here that $M_{XE} = M_{XN}$ and $\lambda_E = \lambda_N$.}
\begin{tabular}{|c|c|c|c|c|c|c|c|c|c||c|c|c||c|c|c|c||c|c|}\cline{1-19}
\multicolumn{19}{|c|}{$\Gamma = 5~ {\rm GeV}$} \\ \hline
$M_{XD}$&$M_{XE}$&$M_{\widetilde{XD}}$&$M_{\widetilde{XE}}$&$\widetilde{M}_{XD}$&$\widetilde{M}_{XE}$&$\lambda_{D}$&$\lambda_E$&$A_{D}$&$A_E$&$\Gamma_{\gamma \gamma}$&$\Gamma_{gg}$&$\Gamma_{XE+XN}$&${\rm Br}_{\gamma \gamma}$&${\rm Br}_{gg}$&${\rm Br}_{XE+XN}$&${\rm Br}_{\rm DM}$&$\sigma_{\gamma \gamma}^{8~{\rm TeV}}$&$\sigma_{\gamma \gamma}^{13~{\rm TeV}}$\\ \hline \hline
$	1200	$&$	255	$&$	2050	$&$	400	$&$	1662	$&$	308	$&$	0.65	$&$	0.351	$&$	3600	$&$	1530	$&$	0.0037	$&$	0.185	$&$	1.45	$&$	0.0007	$&$	0.037	$&$	0.290	$&$	0.672	$&$	0.38	$&$	1.77	$	\\ \hline
$	1350	$&$	225	$&$	1750	$&$	330	$&$	1114	$&$	241	$&$	0.55	$&$	0.351	$&$	4050	$&$	1350	$&$	0.0077	$&$	0.138	$&$	1.88	$&$	0.0015	$&$	0.028	$&$	0.376	$&$	0.594	$&$	0.60	$&$	2.77	$	\\ \hline
$	1200	$&$	265	$&$	1800	$&$	310	$&$	1342	$&$	161	$&$	0.60	$&$	0.351	$&$	3600	$&$	1590	$&$	0.0083	$&$	0.178	$&$	1.30	$&$	0.0017	$&$	0.036	$&$	0.260	$&$	0.702	$&$	0.82	$&$	3.83	$	\\ \hline
$	1400	$&$	235	$&$	1750	$&$	330	$&$	1050	$&$	232	$&$	0.70	$&$	0.351	$&$	4200	$&$	1410	$&$	0.0081	$&$	0.218	$&$	1.74	$&$	0.0016	$&$	0.044	$&$	0.348	$&$	0.607	$&$	0.99	$&$	4.58	$	\\ \hline
$	1050	$&$	225	$&$	1800	$&$	300	$&$	1462	$&$	198	$&$	0.70	$&$	0.351	$&$	3150	$&$	1350	$&$	0.0072	$&$	0.284	$&$	1.88	$&$	0.0014	$&$	0.057	$&$	0.376	$&$	0.565	$&$	1.14	$&$	5.32	$	\\ \hline
$	1000	$&$	215	$&$	1450	$&$	330	$&$	1050	$&$	250	$&$	0.65	$&$	0.351	$&$	3000	$&$	1290	$&$	0.0075	$&$	0.319	$&$	2.02	$&$	0.0015	$&$	0.064	$&$	0.404	$&$	0.530	$&$	1.34	$&$	6.22	$	\\ \hline
$	1000	$&$	255	$&$	1500	$&$	330	$&$	1118	$&$	209	$&$	0.65	$&$	0.351	$&$	3000	$&$	1530	$&$	0.0088	$&$	0.307	$&$	1.45	$&$	0.0018	$&$	0.061	$&$	0.290	$&$	0.647	$&$	1.50	$&$	7.00	$	\\ \hline
\end{tabular}
\label{tab:5GeV}
\end{table}

\begin{table}[htbp]
\centering
\tiny
\caption{Decay widths and production cross-sections for a total decay width of $\Gamma = 45$~GeV for some sample points. All masses and decay widths are in GeV. The cross-sections are in femtobarns (fb). The ${\rm Br}_{\rm DM}$ represents the branching ratio allocated to dark matter. For simplicity, we assume here that $M_{XE} = M_{XN}$ and $\lambda_E = \lambda_N$.}
\begin{tabular}{|c|c|c|c|c|c|c|c|c|c||c|c|c||c|c|c|c||c|c|}\cline{1-19}
\multicolumn{19}{|c|}{$\Gamma = 45~ {\rm GeV}$} \\ \hline
$M_{XD}$&$M_{XE}$&$M_{\widetilde{XD}}$&$M_{\widetilde{XE}}$&$\widetilde{M}_{XD}$&$\widetilde{M}_{XE}$&$\lambda_{D}$&$\lambda_E$&$A_{D}$&$A_E$&$\Gamma_{\gamma \gamma}$&$\Gamma_{gg}$&$\Gamma_{XE+XN}$&${\rm Br}_{\gamma \gamma}$&${\rm Br}_{gg}$&${\rm Br}_{XE+XN}$&${\rm Br}_{\rm DM}$&$\sigma_{\gamma \gamma}^{8~{\rm TeV}}$&$\sigma_{\gamma \gamma}^{13~{\rm TeV}}$\\ \hline \hline
$	1200	$&$	255	$&$	1750	$&$	340	$&$	1274	$&$	225	$&$	0.70	$&$	0.351	$&$	3600	$&$	1530	$&$	0.0090	$&$	0.250	$&$	1.45	$&$	0.00020	$&$	0.0055	$&$	0.032	$&$	0.962	$&$	0.14	$&$	0.65	$	\\ \hline
$	950	$&$	255	$&$	1300	$&$	300	$&$	887	$&$	158	$&$	0.65	$&$	0.351	$&$	2850	$&$	1530	$&$	0.0079	$&$	0.380	$&$	1.45	$&$	0.00018	$&$	0.0085	$&$	0.032	$&$	0.959	$&$	0.19	$&$	0.87	$	\\ \hline
$	1100	$&$	255	$&$	1450	$&$	350	$&$	945	$&$	240	$&$	0.70	$&$	0.351	$&$	3300	$&$	1530	$&$	0.0092	$&$	0.337	$&$	1.45	$&$	0.00021	$&$	0.0075	$&$	0.032	$&$	0.960	$&$	0.19	$&$	0.90	$	\\ \hline
$	1000	$&$	265	$&$	1350	$&$	330	$&$	907	$&$	197	$&$	0.70	$&$	0.351	$&$	3000	$&$	1590	$&$	0.0091	$&$	0.401	$&$	1.30	$&$	0.00020	$&$	0.0089	$&$	0.029	$&$	0.962	$&$	0.23	$&$	1.05	$	\\ \hline
$	900	$&$	265	$&$	1250	$&$	340	$&$	867	$&$	213	$&$	0.65	$&$	0.351	$&$	2700	$&$	1590	$&$	0.0094	$&$	0.420	$&$	1.30	$&$	0.00021	$&$	0.0093	$&$	0.029	$&$	0.962	$&$	0.25	$&$	1.14	$	\\ \hline
$	950	$&$	265	$&$	1300	$&$	350	$&$	887	$&$	229	$&$	0.70	$&$	0.351	$&$	2850	$&$	1590	$&$	0.0097	$&$	0.441	$&$	1.30	$&$	0.00021	$&$	0.0098	$&$	0.029	$&$	0.961	$&$	0.26	$&$	1.23	$	\\ \hline
$	800	$&$	265	$&$	1200	$&$	360	$&$	894	$&$	244	$&$	0.70	$&$	0.351	$&$	2400	$&$	1590	$&$	0.0099	$&$	0.580	$&$	1.30	$&$	0.00022	$&$	0.0129	$&$	0.029	$&$	0.958	$&$	0.36	$&$	1.67	$	\\ \hline
\end{tabular}
\label{tab:45GeV}
\end{table}

The 750~GeV diphoton production cross-sections observed by the CMS collaboration are $\sigma(pp \rightarrow S \rightarrow \gamma \gamma) = 0.5 \pm 0.6~{\rm fb}$ at $\sqrt{s} = 8$~TeV~\cite{Khachatryan:2015qba}and $6 \pm 3~{\rm fb}$ at $\sqrt{s} = 13$~TeV~\cite{bib:CMS_diphoton}, while the ATLAS collaboration observed $\sigma(pp \rightarrow S \rightarrow \gamma \gamma) = 0.4 \pm 0.8~{\rm fb}$ at $\sqrt{s} = 8$~TeV~\cite{Aad:2014ioa} and $10 \pm 3~{\rm fb}$ at $\sqrt{s} = 13$~TeV~\cite{bib:ATLAS_diphoton}. Replicating the strategy of Ref.~\cite{Li:2016xcj}, we constrain the total decay width to $\Gamma \sim 5 - 45$~GeV. To reproduce the observed production cross-sections, we constrain the model using $\Gamma_{\gamma \gamma} \Gamma_{gg} / M_S^2 \gtrsim 10^{-9}$. 

The effective loop-level couplings amongst the Standard Model gauge bosons and scalar $S$ are given by
\begin{equation}
-{\cal L} = \frac{S}{M_S} \left[ \kappa_{EM} F_{\mu \nu}^{EM} F^{EM \smallskip \mu \nu} + \kappa_3 G_{\mu \nu}^a G^{\mu \nu \, a}   \right]
\label{lagrangian}
\end{equation}
where $F_{\mu \nu}^{EM}$ and $G_{\mu \nu}^a$ are the photon and gluon field strength tensors, respectively, with $a = 1,~2,..8$. The effective operators are represented by $\kappa_{EM}$ and  $\kappa_3$, which are written as
\begin{equation}
\kappa_{EM} = \frac{\alpha_{EM}}{4 \pi} \left[ \sum_{f} \frac{\lambda_f M_S}{M_f} Q_f^2 N_{EM}^f F_f + \sum_{\widetilde{f}} \frac{\lambda_f A_f M_S}{M_{\widetilde{f}}^2} Q_{\widetilde{f}}^2 N_{EM}^f F_{\widetilde{f}} \right]
\label{kem}
\end{equation}
\begin{equation}
\kappa_{3} = \frac{\alpha_{3}}{4 \pi} \left[ \sum_{f} \frac{\lambda_f M_S}{M_f} N_{3}^f F_f + \sum_{\widetilde{f}} \frac{\lambda_f A_f M_S}{M_{\widetilde{f}}^2} N_{3}^f F_{\widetilde{f}} \right]
\label{k3}
\end{equation}
where  $N_{EM}^{XD} = 3$, $N_{EM}^{XE} = 1$, and $N_{3}^{XD} = 1$, and the functions $F_f$ and $F_{\widetilde{f}}$ are expressed as
\begin{eqnarray}
&& F_f = 2 \chi \left[ 1 + ( 1- \chi) f(\chi) \right] \\
&& F_{\widetilde{f}} = \chi \left[ -1 + \chi f(\chi) \right] \
\label{feq}
\end{eqnarray}
with the function $\chi$ denoted by
\begin{equation}
\chi = 4  \frac{M_{f/ \widetilde{f}}^2}{M_S^2}
\label{chi}
\end{equation}
The triangle loop functions $f(\chi)$ are defined here as
\begin{eqnarray}
f(\chi)=
	\begin{cases}
		\arcsin^2[\sqrt{\chi^{-1}}]  & \mbox{if } \chi \geq 1 \\
		-\frac{1}{4} \left[ \ln \frac{1 + \sqrt{1 - \chi}}{1 - \sqrt{1 - \chi}} - i \pi \right]^2 & \mbox{if } \chi < 1.
	\end{cases}
\label{loop_function}
\end{eqnarray}
The diphoton and digluon decay widths in ${\cal F}$-$SU(5)$ are computed from
\begin{equation}
\Gamma_{\gamma \gamma} = \frac{\left| \kappa_{EM} \right|^2}{4 \pi} M_S
\label{Gamma_pp}
\end{equation}
\begin{equation}
\Gamma_{gg} = \frac{2 \left| \kappa_{3} \right|^2}{\pi} M_S
\label{Gamma_gg}
\end{equation}

\noindent The diphoton production cross-section is calculated from
\begin{equation}
\sigma(pp \rightarrow S \rightarrow \gamma \gamma) = K\frac{C_{gg} \Gamma(S \rightarrow gg) \Gamma(S \rightarrow \gamma \gamma)}{s \Gamma M_S}
\label{xsection_pp}
\end{equation}

\noindent where K is the QCD K-factor, $\Gamma$ is the total decay width, $\sqrt{s}$ is the proton-proton center of mass energy, and $C_{gg}$ is the dimensionless partonic integral computed for an $M_S = 750$~GeV resonance, yielding $C_{gg} = 174$ at $\sqrt{s} = 8$~TeV and $C_{gg} = 2137$ at $\sqrt{s} = 13$~TeV~\cite{Franceschini:2015kwy}. We use the gluon fusion K-factor of 1.98.

We construct our intersecting D-brane model with the ($XD$,$XD^c$) and ($XL$,$XL^c$) vector-like particles, implementing three copies of the ($\bf{5}$,$\overline{\bf 5}$). For the calculations, we decompose the $(XL, XL^c)$ multiplet into its $(XE, XE^c)$ and $(XN, XN^c)$ components. Given the null $XN$ electric charge $Q_{XN} = 0$, no constraints can be placed on $M_{XN}$, $\lambda_N$, or $A_N$ in the model via the production cross-section calculations, so for simplicity we set $M_{XN} = M_{XE}$ and $\lambda_N = \lambda_E$ when computing the decay of the scalar $S$ directly to the $XN$ multiplet. The multiplets $(XD, XD^c)$ and $(XE, XE^c)$ participate in the $S \rightarrow \gamma \gamma$ loop diagrams and only $(XD, XD^c)$ in the $S \rightarrow gg$ loops, hence there are 8 free parameters in the effective operators $\kappa_{EM}$ and $\kappa_3$ consisting of the Yukawa couplings $\lambda_f$, trilinear A term couplings $A_f$, fermionic component masses $M_f$, and scalar component masses $M_{\widetilde{f}}$. In total, there are 10 parameters to compute:
\begin{equation}
M_{XD},~M_{XE},~M_{\widetilde{XD}},~M_{\widetilde{XE}},~\widetilde{M}_{XD},~\widetilde{M}_{XE},~\lambda_D,~\lambda_E,~A_D,~A_E
\label{list}
\end{equation}
though the supersymmetry breaking soft terms $~\widetilde{M}_{XD}$ and  $~\widetilde{M}_{XE}$ can be trivially computed from the fermionic and scalar components using the following relations
\begin{eqnarray}
&& M_{\widetilde{XD}}^2 = M_{XD}^2 + \widetilde{M}_{XD}^2 \\
&& M_{\widetilde{XE}}^2 = M_{XE}^2 + \widetilde{M}_{XE}^2 \
\label{soft_eq}
\end{eqnarray}

The freedom on the 8 free-parameters engenders a large D-brane model parameter space. We treat the fermionic component of the vector-like particle and its soft supersymmetry breaking term independently, such that $M_f \neq \widetilde{M}_f$. Recent constraints at the LHC on vector-like $B$ quarks~\cite{atlas-vector-like} provide lower limits of around 735~GeV for the $XD$ multiplet, allowing a light $XE$ multiplet, which contributes to invisible branching fractions when $M_{XE} < 375$~GeV.
For the large Yukawa couplings $\lambda_D$ and $\lambda_E$ at the unification scale,
there exist the quasi focus points with $\lambda_D\simeq 0.70$ and $\lambda_E\simeq 0.351$
at low energy, respectively~\cite{Dutta:2016jqn}. 
Constraints are placed on the $A$ terms of $A_{D} \lesssim 3 M_{D}$ and $A_{E} \lesssim 6 M_{E}$ to preclude premature breaking of the $SU(3)_C \times U(1)_{EM}$ gauge symmetry. We provide some sample benchmark points in TABLE~\ref{tab:5GeV} and TABLE~\ref{tab:45GeV} to illustrate that the model can accommodate the observed diphoton cross-section.  Note that the cross-sections $\sigma_{\gamma \gamma}^{8~{\rm TeV}}$ and $\sigma_{\gamma \gamma}^{13~{\rm TeV}}$ in TABLES~\ref{tab:5GeV} - \ref{tab:45GeV} display a gain of 4.65 from 8~TeV to 13~TeV. We use values of the coupling constants at the $M_Z$ scale in our calculations of $\alpha_3 = 0.1185$ and $\alpha_{EM} = 128.91^{-1}$. 

The diphoton and digluon decay modes are only a small portion of the total decay width, as shown in TABLES~\ref{tab:5GeV} - \ref{tab:45GeV}. These practical constraints on the diphoton and digluon decay widths allocate the majority of the total width to the other decay channels, for instance, to dark matter invisibly and/or $S \rightarrow XE  XE^c$ and $S \rightarrow XN  XN$, kinematically allowed for both $M_{XE}$ and $M_{XN}$ less than 375~GeV, which is reflected in the $\Gamma_{XE+XN}$ decay width and $Br_{XE+XN}$ branching ratio in TABLES~\ref{tab:5GeV} - \ref{tab:45GeV}. Specifically, the decay width for the decay into pairs of fermions $S \rightarrow f  \overline{f}$ is given by
\begin{equation}
\Gamma(S\rightarrow f \overline{f}) = \frac{1}{16\pi} M_S \lambda_{f}^2 \left(1-\frac{4M_f^2}{M_S^2}\right)^{3/2}.
\end{equation}
where we take $M_{XE} = M_{XN}$ and $\lambda_E = \lambda_N$ in the calculations.

\section{Conclusion}

We have studied the diphoton excesses near $750$~GeV recently reported by the ATLAS and CMS
collaborations within the context of a phenomenologically interesting 
intersecting/magnetized D-brane model on a toroidal orientifold.  
We have shown that the model contains a SM singlet scalar as well as vector-like
quarks and leptons.  In addition, we have shown that the singlet scalar has Yukawa couplings
with vector-like quarks and leptons such that it may be produced in proton-proton
collisions via gluon fusion as well as decay to diphotons through loops
involving the vector-like quarks.  Moreover, the required vector-like quarks and leptons
may appear in complete $SU(5)$ multiplets so that gauge coupling unification may be 
maintained. In particuar, we showed that we may have three copies of 
$\mathbf{5}+\mathbf{\overline{5}}$ representations of $SU(5)$ in the light spectrum which are 
present in the model. Finally, we showed that the diphoton excesses observed by the 
ATLAS and CMS collaborations may be accommodated.  

It should be emphasized that we have obtained these results within the context of a 
complete, globally consistent string model.  This particular model has many interesting 
phenomenological features such as automatic gauge coupling unification, realistic 
Yukawa mass matrices for quarks and leptons, and mimimal exotic matter.  It is of note
that the singlet fields required to explain the diphoton signal arise from this extra matter.
In addition, the required vector-like quarks are naturally present in the spectrum.  Finally,
the Yukawa couplings between the singlet fields and the vector-like quarks are allowed
by the global symmetries arising from $U(1)$ factors whose gauge bosons become heavy at the
string scale via the Green-Schwarz mechanism, a result which is completely non-trivial.  

An interesting question is whether or not is it possible to obtain supersymmetry partner
spectra from the model which take into account the light vector-like matter.  As mentioned
earlier, the vector-like quarks and leptons have Yukawa couplings with the Higgs fields in the
model, and thus may raise the Higgs mass by up to a few GeV.  This may alleviate the problem 
with electroweak fine-tuning.  In addition, the vector-like matter affects the RGE running 
of the soft masses.  Finally, the soft supersymmetry breaking masses may be calculated at the 
string scale in the model.  Thus, it would be very interesting to study the possible 
supersymmetry partner spectra obtainable in the model including the extra vector-like matter. 
We plan to study this in future work.

\begin{acknowledgments}

This research was supported in part by the Natural Science Foundation of China
under grant numbers 11135003, 11275246, and 11475238 (TL), and 
by the DOE grant DE-FG02-13ER42020 (DVN).

\end{acknowledgments}

\end{document}